\documentclass[twocolumn,floats,prl,aps]{revtex4}

\usepackage{graphicx}

\newcommand\beq{\begin{equation}}
\newcommand\eeq{\end{equation}}
\newcommand\bea{\begin{eqnarray}}
\newcommand\eea{\end{eqnarray}}
\newcommand{\nonum}{\nonumber}

\begin{document}

\title{\bf Quantum Phase Transition in Coupled Superconducting Quantum 
Dots Array with Charge Frustration}

\author{\bf Sujit Sarkar}
\address{\it
PoornaPrajna Institute of Scientific Research,
4 Sadashivanagar, Bangalore 5600 80, India.\\
}
\date{\today}


\date{\today}

\begin{abstract}
We present the quantum phase transition in two capacitively coupled arrays
of superconducting quantum dots (SQD). We consider the presence of gate voltage
in each superconducting island. We show explicitly that the co-tunneling
process involves with two coupled SQD arrays, 
near the maximum charge frustration line is not sufficient
to explain the correct quantum phases with physically consistent phase
boundaries. We consider another extra co-tunneling process along each chain to
explain the correct quantum phases with physically
consistent phase boundaries. There is no evidence of supersolid
phase in our study. We use Bethe-ansatz and Abelian bosonization method to 
solve
the problem.\\
\vskip .4 true cm
\end{abstract}
\maketitle


\noindent
Josephson junction arrays have attracted considerable interest in the recent
years owing to their interesting physical 
properties \cite{sondhi,fazi,geer,van,dona,lar,sar,sar2,bal,g1,g2,
gia2,lee,fish}. 
Currently such arrays can
be fabricated in restricted geometries both in one and two dimensions with well
controlled parameters \cite{kuo,havi1,havi2,havi3}. 
At the same time this is one of the 
paradigms to study the physics of quantum
phase transition \cite{sondhi,sar,sar2}. 
The Coulomb charging energy of the system is due to the small
capacitance of the grains or junction that dominate strong quantum phase fluctuations
of the superconducting order parameter and may drive the system into the insulating
state at zero or very low temperature. The quantum fluctuations are controlled
by the parameters of the system such as charging energy of the superconducting
quantum dots (SQD) and Josephson coupling (${E_J}$ ) between them.\\
In two dimensions, the universality class of these transitions has already been
investigated in detail \cite{g1,g2,gia2,lee,fish}. 
Nevertheless there are also quantum phase transition
scenario that can take place in a Josephson junction ladder system as shown in the
Fig. 1, where we consider two capacitively coupled one dimensional SQD 
arrays. 
In this simplest two-leg ladder system, we will be able to predict different
quantum phases due to the interplay between Coulomb charging energy and 
$ E_J $. Here we also show explicitly the role of co-tunneling effect.
For large Coulomb charging energy sequential tunneling is not energetically
favored process in the system. The major tunneling process 
in the system occurs via the co-tunneling process \cite{bruu}. Its origin is quantum
mechanical. In this process, tunneling occurs through the virtual state with
energy equal to the on-site Coulomb charging energy. An appropriate co-tunneling
process extract the correct quantum phases in the system.  
Otherwise, it leads to the
wrong analysis of the system as we have found in Ref.\cite{choi}. In our previous
studies we only emphasis the one dimensional SQD array, we only explain the
importance of next-nearest-neighbor (NNN) $E_J $ \cite{sar,sar2} 
.\\
In this paper we study the quantum phase transition of capacitively coupled two
arrays of SQD and they are connected through Josephson
junctions. We consider the presence of gate voltage in each SQD which introduce
the charge frustration in the system. 
At the charge frustration line the Coulomb charging energy of the system
is degenerate for the difference of one Cooper pair
in the island. Its also refer in the literature as a charge degeneracy point.
Our main motivation is to study the different
quantum phases around the maximum charge frustration line. 
Choi $et~al.$ \cite{choi} have also studied the same problem 
They have shown that in a coupled
chain the major transport along both the chains occurs via co-tunneling of
the electron-hole pairs
They have not done any rigorous analytical exercise based on techniques,
which are applicable in low dimensional quantum many
particle systems and that seed mistakes.
There are three major mistakes
in that paper
(1) There are two kinds of Luttinger liquid (LL) phase (describe as RL1 and RL2)
but they have obtained only one.
(2) Incorrect quantum phase analysis and physically
inconsistent phase boundaries due to lack of correct analytical derivations
and physical interpretations.
(3). There is no evidence of super solid (SS) phase in this
model.\\
We show explicitly in
this study that a single co-tunneling process in the two leg SQD ladder system
is not sufficient to produce the correct quantum phases with physically consistent
phase boundaries. We also show that the co-tunneling process along each chain is
also necessary for the correct phase boundaries.
In our model system, superconducting islands are connected with 
$E_J $. The charging energies along each array are the on-site Coulomb
charging energy due to the self capacitance, $ {E_0} = \frac{e^2}{2 C_0 }$ and the
junction charging energy, ${E_1} = \frac{e^2}{2 C_1 }$ 
due to junction capacitance $ C_1 $. The arrays are coupled
to each other via the capacitance $C_I $, it contributes charging energy,
$ {E_I} = \frac{e^2 }{2 C_I } $. But there are no Josephson coupling between the
two arrays. The array capacitances are small, which leads to 
$ {E_J} << {E_0},{E_1}$. Therefore each arrays are in the insulating phase in the
absence of co-tunneling effect. Here we consider the limit 
${E_I} << {E_0}, {E_1}$. The applied gate voltage induce a charge in every
superconducting island by an amount, $ {n_g} = \frac{C_0 V_g}{2e} $, this applied
gate voltage introduce the charge frustration in the system. We do the quantum
phase analysis near the maximum charge frustration line ($|{n_g} - N -1/2 | << 1$),
around this line we obtain several interesting quantum phases 
which have not been noticed
in the previous studies \cite{dona,lar,choi}. 
\begin{figure}
\includegraphics[scale=0.50,angle=0]{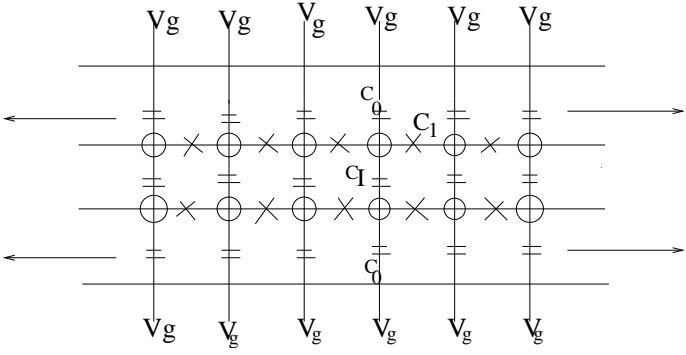}
\caption{This is the schematic diagram of our model systems. Two arrays
of superconducting quantum dots are connected through the capacitance, $C_I $.
${C_0}$ is the self capacitance of the dot and $C_1 $ is the junction capacitance.
In our superconducting circuit, small circles denote the dots and crosses denote
the Josephson junctions. ${V_g}$ is the gate voltage applied in each dot. }
\label{Fig. 1 }
\end{figure} 
The Hamiltonian of the system is given by
\bea
H & = & 2 {e^{2}} ~ \sum_{l l^{'}, x x^{'}} ( {n_l}(x) - {n_g} )~ 
{C}_{l, l^{'}}^{-1} (x, {x^{'}} )
~( {n_l}(x^{'} ) - {n_g} ) \nonum\\
& & - {E_J} \sum_{l,x} cos ({{\theta}_l} (x) - {{\theta}_{l} (x+1) })
\eea
$n_{l} (x)$ is the number of Cooper pairs and ${{\theta}_l} (x)$ is the phase of
the superconducting order parameter at the site x of the lth array. $n_{l} (x)$
and ${{\theta}_l } (x)$ are the quantum mechanically conjugate variables. One can
write the capacitance matrix in the block form.
\beq
{C}_{l, l^{'}}^{-1} (x, {x^{'}} ) = {C} (x, {x^{'}} ) 
A + {\delta}_{x, x^{'}} { C_{I} } B
\eeq
Where 
 \[ A=  \left (\begin{array}{cc}
      1 & 0 \\
      0 & 1 \\
        \end{array} \right ) \], 
\[ B= \left (\begin{array}{cc}
      1 & -1 \\
      -1 & 1 \\
        \end{array} \right ) \] 
\beq
C(x, x^{'}) ~=~ C_{0} {\delta}_{x, x^{'}} ~+ ~ 
{C_1} [2 {\delta}_{x, x^{'}} - {\delta}_{x, {x^{'}+1}}
- {\delta}_{x, {x^{'}-1}}] 
\eeq
\beq
H~=~ {H_c }^{(0)} ~+~ {H_c }^{(1)}+{H_J }
\eeq
with the components 
\beq
{H_c }^{(0)} = U_{0} \sum_{x} {[ {n_{+}} (x) - 2 {n_g}]}^{2}
+ {V_0} \sum_{x} ({n_{-}} (x))^{2}
\eeq
\bea
{H_c }^{(1)} & = & U_{1} \sum_{x} [ {n_{+}} (x) - 2 {n_g}]
[ {n_{+}} (x+1) - 2 {n_g}] \nonum \\
& & + {V_1} \sum_{x} {n_{-}} (x) {n_{-}} (x+1)
\eea
\beq
{H_J} =  - {E_J} \sum_{l,x} cos ({{\theta}_l} (x) - {{\theta}_{l} (x+1) })
\eeq
Where $ n_{\pm} (x) = n_{1} (x) ~\pm~ n_{2} (x)$ and the coupling strengths are given
by $ U \sim 2 {E_0} $, $ {U_1} = 4 \frac{C_1}{C_0} $, ${V_0} \sim {E_I}$
and ${V_1} \sim \frac{{C_1}{C_I}} {E_I}$.
The first term of the Hamiltonian $ {H_c }^{(0)}$ contains an important message
regarding the difference between the charging energy at near to the maximum 
charge frustration line and the particle-hole symmetric line (${n_g}=0$). The
charge configurations which do not satisfy the
condition for ${n}_{+} (x) = 1 $ at the maximum charge frustration line generates a
gap in the excitation spectrum of the order of on site charging energy.
The ground state of $ {H_c }^{(0)}$ separated from the excited state by the gap
of the order of $E_I $. This excited state has two fold degeneracy for every values
of x, corresponding to $ n_{-} (x)= \pm 1$. 
It is convenient to work in the charge configuration with $n_{+} (x) =0 $
and $n_{-} (x)= \pm 1$ is termed as a reduce Hilbert space of the problem.
Presence of finite $E_j $ lifts this
degeneracy and the ground state of ${H_c}^{0} $ is mixed with the state 
${n_{-} (x)} = {\pm 2} $. Now the relevant reduce Hilbert space, 
${n}_{+} (x) = 0 $  and 
${n}_{-} (x) = 0, {\pm 2} $. Here we mention the effective model of our system,
wherein we follow the references 
. They have found the effective
Hamiltonian up to the second order in $\frac{E_J}{E_0}$. 
\beq
 H_{eff} = P [ H + {H_J} \frac{1 - P}{E - {H_C}^{0}} {H_J} ] P 
\eeq
Here $P$ is the projection operator onto the reduce space and finally
they have obtained the effective Hamiltonian.
\bea
H_{eff} & = & {\gamma} J \sum_{x} {S^z } (x) {S^z } (x+1) \nonum\\
& & \frac{-J}{2} \sum_{x} [{S^{+} (x) S^{-} (x+1) + h.c}]
\eea  
where the $XY$ component of exchange interaction and the z component of
exchange anisotropy are respectively 
$ J = \frac{{E_J}^2 }{ 4 E_{0}} $ and 
$\gamma =\frac{16 {\lambda}^{2} {E_I}^{2}}{ {E_J}^2 } $. 
No analytical expression for ${\lambda}^2 $ is given in Ref. .
Here we estimate ${{\lambda}^2} = \frac{ {C_1} {E_0}}{ {C_I} {E_I}} $.
Pseudo spin
operators of the above Hamiltonian are the following:
\bea
{S^z} (x) & = & P \frac{ {n_1}(x) - {n_2}(x)}{2} P \nonum\\
{S^{+}} (x) & = & P e^{-i {\theta}_1  (x)} (1- P) e^{-i {\theta}_2 (x)} P \nonum\\
{S^{-}} (x) & = & P e^{-i {\theta}_2 (x)} (1- P) e^{-i {\theta}_1 (x)} P. 
\eea 
Now we would like to explain the pseudo-spin terms of the Hamiltonian in terms of
charge representation from where we start. The first term of the Hamiltonian 
originates from the first term of Eq. 9 which the Coulomb charging energy of the
SQD in the form of $n{(x)}$. The second term of the Hamiltonian originates
from the co-tunneling process, i.e, from the second term of Eq. 9.
They have obtained the Hamiltonian in an effective one dimensional Hamiltonian
but they have not done any correct analytical calculations based on the 
Abelian bosonization and  Bethe ansatz methods which are suitable for one
dimensional quantum many body system. Here we would like to do the 
quantum phase boundaries analysis based on the exact Bethe ansatz calculation
based results. We will present the short comings of this effective Hamiltonian
from that analysis.\\
\bea
S^{z} (x) & = &  \psi^{\dagger} (x) \psi (x) - 1/2 ~,\nonum\\
S^{-} (x) & = & { \psi (x)} ~\exp [i \pi \sum_{j=-\infty}^{x-1} n_j]~, \nonum\\
S^{+} (x) & = & { \psi^{\dagger}} (x)~\exp [-i \pi \sum_{j=-\infty}^{x-1} n_j]~,
\eea
where $n (x) = {\psi^{\dagger}} (x) {\psi (x)} $ is the fermion number at 
the site $x$.
\begin{center}
\bea
 {H}_{eff} & = & - {\gamma} {E_{J}} ~\sum_x ~({\psi}^{\dagger}(x) \psi{(x)} - 1/2)\nonum\\
& & ({\psi}^{\dagger} (x+1) \psi{(x+1)} - 1/2) \nonum\\
& & \frac{-J}{2} ~\sum_x ~( {\psi}^{\dagger} (x+1) \psi{(x)} + {\rm h.c.})
\eea
\end{center}
In order to study the continuum field theory of these Hamiltonians,
we recast the spinless
fermion operators in terms of field operators by this relation \cite{gia1}
$ {\psi}(x)~=~~[e^{i k_F x} ~ {\psi}_{R}(x)~+~e^{-i k_F x} ~ {\psi}_{L}(x)] $
where ${\psi}_{R} (x)$ and ${\psi}_{L}(x) $ 
describe the second-quantized fields of right- and
the left-moving fermions respectively.
We want to express the fermionic fields in terms of bosonic fields by the relation
$ {{\psi}_{r}} (x)~=~~\frac{U_r}{\sqrt{2 \pi \alpha}}~
~e^{-i ~(r \phi (x)~-~ \theta (x))}$ ,
where $r$ is denoting the chirality of the fermionic fields,
right (1) or left movers (-1).
The operators $U_r$ commute with the bosonic field. $U_r$ of different species
commute and $U_r$ of the same species anti commute. $\phi$ field corresponds to the
quantum (bosonic) fluctuations of spin and $\theta$ is the dual field of $\phi$.
They are
related by the relation 
$ {\phi}_{R}~=~~ \theta ~-~ \phi$ and  $ {\phi}_{L}~=~~ \theta ~+~ \phi$.
After doing the continuum field theory exercise, $ H_{eff} $
become
\beq
 H_{eff} = H_{0} + \frac{\gamma J}{2 \pi {\alpha}^2 } \int ~dx
: cos (4 \sqrt{K_1} \phi (x) ): 
\eeq
$ {H_0}= \frac{v}{2 \pi} \int dx
[ {({{\partial}_x} {\theta})}^2 + {({{\partial}_x} {\phi})}^2 ] $
is the non-interacting part of the Hamiltonian.
The collective velocity
of the system ($v$) and $K_1 $ are the two LL parameter.
We use exact Bethe ansatz solution to calculate
\beq
 {K_1}~=~ \frac{\pi}{\pi~+~ 2 sin^{-1}{\gamma } } 
\eeq
\begin{figure}
\includegraphics[scale=0.45,angle=0]{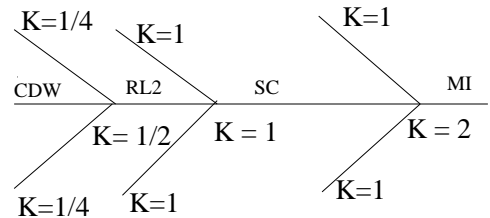}
\caption{We present the one dimensional schematic diagram for
quantum phases as a function of Luttinger liquid parameters ($K$).
In the figure, we present also values of K at the phase boundaries.
The value of $K$ at the wings of
the arrows denote the value of $K$ at the phase boundaries.
CDW is the charge density wave phase. RL2 is the repulsive Luttinger
liquid phase of second kind. SC is the superconducting phase. The
limit $K =2$ only achieve when the system is in the first order
commensurability, i.e., each dots are occupied with integer
number of Cooper pairs. MI is the Mott-insulating phase of the system. 
The presence of gate voltage is not explicit in K, therefore the
RL1 phase not appears in figure, we can only achieve this phase when the
applied gate voltage exceed the CDW gap.  
}
\label{Fig. 2 }
\end{figure}
For $ J < {\gamma} J $ with relatively small applied gate voltage,
the antiferromagnetic Ising interaction dominates the physics of anisotropic
Heisenberg chain. When the applied gate voltage is large the chain is in the
ferromagnetic state. In the language of interacting Cooper pairs, the Neel
phase is the Charge density wave phase with period 2 i.e there is only one
Cooper pair in every two sites. For the ferromagnetic phase system is in the
Mott insulating phase.
The emergence of two LL phases can be ascribed to the following
reasons: RL1 and RL2, respectively occur due to the
commensurate-incommensurate transition and the criticality of the Heisenberg
XY model. The system is either
in RL1 for $K < 1/2 $ or in RL2 for $K > 1/2 $. The physical significance of
RL1 is that the coupling term is relevant but the applied gate voltage on
SC island breaks the CDW gapped phase, whereas in RL2, system is gapless for
$K_1 > 1/2 $.\\
In Fig. 2, we present quantum phases of our study as a function of $K$. 
We apply the Luther-Emery (LE) trick \cite{luth} in the massive phase of sine-Gordon
field theory to evaluate $K_1 $ (=1/4) at the phase boundary between the
CDW and RL1 and also for CDW and RL2. 
RL1 phase is not explicit in the figure because we only achieve
this phase under the application of gate voltage when it exceed the gap of
CDW state.
The value of $K_1 $ at the 
phase boundary between RL2 and CDW phase is $1/4$. From the analysis of $K_1 $
we obtain $ 16 {{\lambda}^2} {E_I}^{2} = - {E_J}^2 $, this condition is
unphysical because $\lambda$, $E_I $ and $E_J $ are (+ve) quantities.
The value of $K$ is $ 1 $ at the phase boundary between the RL2 and superconducting
phase. The analysis yields the condition $ 16 {\lambda}^{2} {E_I}^{2} =0 $, which is
again unphysical. There is no chance to get superconductivity 
in this model Hamiltonian. We surprise that without doing any correct
quantum analysis of this  
model Hamiltonian, the authors of Ref. \cite{choi} have claimed the existence of 
SC. 
\\
To get the correct physical behavior between the different quantum phases,
one has to be considered the Co-tunneling processes along each chain separately.
We have realized during our calculations that to get an
attractive interaction between the Jordan-Wigner (spinless)
fermions, we have to consider the higher order expansion
in $\frac{E_{J1}}{E_{C0}}$. This higher order expansion leads
to the virtual state with energies exceeding $E_{C0}$. In this
second order process, 
extra contribution appears as 
\bea
 H_{extra} & = & \frac{-3 {E_J}^{2}}{4 E_0} \sum_{x} S^{z} (x) S^{z} (x+1) \nonum\\
& & - \frac{{E_J}^{2}}{ E_0} \sum_{x} ( S^{\dagger} (x+2) S^{-} (x) + h.c )
\eea
\cite{lar,sar,sar2,giu1}.
The total effective Hamiltonian of the system under the combined
co-tunneling process is
\bea
H_{eff} & = & ({\gamma} J - \frac{3 {E_J}^{2}}{4 E_0})
\sum_{x} S^{z} (x) S^{z} (x+1)
\nonum\\
& & - \frac{J}{2} \sum_{x} ( {S^{\dagger}} (x) S^{-} (x+1) + h.c ) \nonum\\
& & - \frac{{E_J}^{2}}{ E_0} \sum_{x} ( S^{\dagger} (x+2) S^{-} (x) + h.c )
\eea
After doing the quantum field theory, we get the effective Hamiltonian
\beq
{H_{eff}} = {H_0 } + \frac{\gamma J - \frac{3 {{E_J}^2 } }{4 E_0 } }
{2 \pi {\alpha}^2 } \int ~dx
: cos (4 \sqrt{K_2} \phi (x) ):
\eeq
LL parameter of $H_{eff}$ is
\beq
{K_2 }= \sqrt{[\frac{J}{J + {4/\pi}( \gamma J - \frac{3 {{E_J}^2 } }{4 E_0 })} ]}   
\eeq
As we notice from the Fig. 2 that value of 
${K_2} =1/4$ at the phase boundary between RL2 and CDW phase. The
parametric condition at the phase boundary is the following
$ \frac{16 {\lambda}^{2} {{E_I}^2} }{ {E_J}^{2}} = \frac{15 \pi}{4} +3 $  
This condition is consistent physically. Therefore we prove the importance
of this extra co-tunneling process. We see from the Fig. 2 that value of the
$K_2 $ is $1$ at the phase boundary between the RL2 and superconducting phase.
The parametric analysis for this value of $K_2 $ yields
$ {16 {\lambda}^{2} {{E_I}^2} } = 3 { {E_J}^{2}} $
, we obtain,
$ {E_J}^{2} = 8 {\lambda}^{2} {E_I}^2 $. This is a physically realizable
condition for phase boundaries.
It is clear from the Fig. 2 that there is no simultaneous presence of
CDW phase and SC phase. SC phase occurs only when $ K >1 $ and
CDW phase occurs only when $1/2 < K<1 $. Therefore there is no
evidence of SS phase for this model system.\\
Conclusions: We have found the all correct quantum phases of this model
Hamiltonian based on the rigorous analytical calculations. 
We obtain the condition for physically consistent phase boundaries which were
absent in the previous studies. There is no evidence of super solid phase.\\
Acknowledgment: The author would like to acknowledge CCMT of the Physics
Department of IISc for extended facility and also Prof. R. Srikanth for
reading the manuscript.


\begin{thebibliography}{0}

\bibitem{sondhi} S. L. Sondhi $et~al.$, Rev. Mod. Phys. {\bf 69}, 315 (1997).

\bibitem{fazi} R. Fazio, and  H. van der Zant, Physics Report
{\bf 355}, 235 (2001).

\bibitem{geer} L. J. Geerligs $et~al.$, Phys. Rev. Lett. {\bf 63}, 326 (1989).

\bibitem{van} H. S. J. van der Zant $et~al.$, Euro. Phys. Lett. {\bf 119}, 
541 (1992).

\bibitem{dona} R. M. Bradley, and S. Doniach, Phys. Rev. B {\bf 30},
1138 (1984).

\bibitem{lar} L. I. Glazmann, and A. I. Larkin, Phys. Rev. Lett. {\bf 39},
3786 (1997).

\bibitem{sar} S. Sarkar, Euro. Phys. Lett, {\bf 71} 980 (2005).

\bibitem{sar2} S. Sarkar, Phys. Rev. B {\bf 75}, 014528 (2007).

\bibitem{bal} R. Baltin, and K. H. Wagenblast, Euro. Phys. Lett,
{\bf 39}, 7 (1997).

\bibitem{g1} E. Granto and J. M. Kosterlitz, Phys. Rev. Lett. 
{\bf 65}, 1267 (1990).

\bibitem{g2} E. Granto, Phys. Rev. B {\bf 42}, 4797 (1990).

\bibitem{gia2} E. Orignac and T. Giamarchi, Phys. Rev. B {\bf 64},
144515 (2001).

\bibitem{lee} H. Lee and M. C. Cha, Phys. Rev. B {\bf 65}, 172505 (2002).

\bibitem{fish} M. P. A. Fisher, G. Grinstein and S. M. Girvin,
Phys. Rev. Lett {\bf 64}, 587 (1990). 

\bibitem{kuo} W. Kuo and C. D. Chen, Phys. Rev. Lett. {\bf 87}, 186804 (2001).

\bibitem{havi1} H. M. Jaeger $et~al.$
Phys. Rev. B {\bf 40}, 182 (1989).

\bibitem{havi2} C. D. Chen $et~al.$, 
Phys. Rev. B {\bf 51}, 15645 (1995).

\bibitem{havi3} E. Chow, P. Delsing, and  D. B. Haviland, Phys. Rev. Lett.
{\bf 81}, 204 (1998).

\bibitem{bruu} H. Bruus, and K. Flensberg, in {\it Many Body Quantum Theory
in Condensed Matter Physics} (Oxford University Press, New York 2004).

\bibitem{choi} Mahn-Soo Choi $et~al.$, Phys. Rev. Lett {\bf 81}, 4240 (1998). 

\bibitem{gia1} T. Giamarchi, in {\it Quantum Physics in One Dimension}
(Claredon Press, Oxford 2004).

\bibitem{luth} We follow the Luther-Emery trick 
during the analysis.
One can write the sine-Gordon Hamiltonian for arbitrary
commensurability as
$ H_2 ~=~ H_{01} ~+~ \lambda \int~ dx~cos (2 n~ \sqrt{K}~\phi (x) ) $,
where $n$ is the
commensurability and $\lambda$ is the coupling strength.
$H_{01}$ is the free part of the Hamiltonian.
We know that for the spinless fermions,
${{\psi}_R}^{\dagger}{{\psi}_L}~+~{{\psi}_L}^{\dagger}{{\psi}_R}
~=~\frac{1}{{2 \pi a}^2} \int dx cos (2 \sqrt{K} \phi (x)) $, which is
similar to the analytical expression of sine-Gordon coupling term
but with the wrong coefficient inside the cosine. One can set
${\tilde {\phi} (x)}~=~ 2 \sqrt{K} \phi (x)$ then the above equation become
$ H_3 ~=~ H_{01} + \lambda \int~dx cos(2 {\tilde {\phi} (x)}). $
$K$ and $\tilde{K}$ are related by the relation, $ K~=~\frac{\tilde{K}}{n^2}$.
At the phase boundary, $\tilde{K}
~=1$ that implies $K=1/n^2$. 

\bibitem{giu1} D. Giuliano, and P. Sodano, Nucl. Phys. B {\bf 711}, 480 (2005).


\end{thebibliography}
\end{document}